\documentclass[aps,prb,preprint,showpacs,showkeys]{revtex4-1}

\usepackage{mathptmx}
\usepackage{graphicx}

\usepackage{graphicx}
\usepackage{epsfig}

\usepackage{amssymb}

\begin{document}

\title{Confirmation in graphene  of  wave packet multilooped dynamics related to fractional quantum Hall state}

\author{J. Jacak, L. Jacak}
\affiliation{Institute of Physics, Wroc{\l}aw University of Technology, Wyb. Wyspia{\'n}skiego 27, 50-370 Wroc{\l}aw, Poland}

\begin{abstract}
Cyclotron braid subgroups are defined in order to identify the topological origin of Laughlin correlations in 2D Hall systems. Flux-tubes and vortices for composite fermion constructions are explained in terms of unavoidably multilooped  cyclotron braids. A  link of braid picture with quasiclassical quantum dynamics is conjectured in order to support the phenomenological model of composite fermions with auxiliary flux-tubes, for Landau level fillings out of $\frac{1}{p}$, $p$ odd.  The even denominator fractional lowest Landau level fillings, including Hall metal at $\nu=\frac{1}{2}$, are also discussed in cyclotron braid terms.  The topological arguments are utilized  to explain novel experimentally observed  features of the fractional quantum Hall state in graphene including   the  triggering role of carriers mobility for this collective state.

\end{abstract}


\maketitle


\section{Introduction}
In order to describe correlations in 2D charged multi-particle systems in the presence of strong perpendicular magnetic field, the famous Laughlin wave-function (LF) was introduced \cite{laughlin2}. The representation of the Coulomb interaction in terms of the so-called Haldane pseudopotential allowed for an observation \cite{haldane,prange,laughlin1} that the LF exactly describes the ground state for $N$ charged 2D particles at the fractional Landau level (LL) filling $1/q$, $q$--odd integer, if one neglects the long-distance part of the Coulomb interaction expressed by a projection on the relative angular momenta of particle pairs for values greater than $q-2$. The Laughlin correlations were next effectively modeled by composite fermions \cite{jain} (CFs)  in terms of auxiliary flux-tubes attached to particles. By virtue of the Aharonov-Bohm effect, the flux-tubes attached to particles produce the required by LF statistical phase shift when particle interchange. The great advantage of the CF construction was recognized in possibility of interpretation of a fractional quantum Hall effect (FQHE) in an external magnetic field as an integer quantum Hall effect (IQHE) in resultant field screened by averaged field of the fictitious flux-tubes \cite{jain}. This allowed for recovery of the main line of FQHE filling factor hierarchy, $\nu=\frac{n}{(p-1)n\pm 1}$, ($p$--odd integer, $n$--integer) \cite{jain}, corresponding to complete filling of $n$ LLs in the screened field assuming that resultant field can be oriented along or oppositely to the external field (thus $\pm$ in the obtained hierarchy). Despite of a wide practical usage of CFs in description of 2D Hall systems, the origin and nature of attached to particle flux-tubes are unclear, similarly as unclear is also the heuristic assumption that the resultant field screened by the mean field of local fluxes can be oriented oppositely to the external field (allowing, in that manner, for the sign minus in the hierarchy formula obtained by mapping of FQHE onto IQHE).

The competitive construction of CFs was also formulated utilizing so-called vortices \cite{vor3,vora1}, collective fluid-type objects (in analogy of vortices in superfluid systems) that are assumed to be pinned to bare fermions and reproducing Laughlin correlations \cite{vor3}.  The vorticity number is, however, assumed without any argumentation, but only in agreement with in advance known LF, and equal to the exponent of the Jastrow polynomial being a factor of the LF. Thus both types of composite particles, with vortices or with flux tubes, are phenomenological in nature, and the question arises as to what is a more fundamental reason of Laughlin correlations  in 2D charged systems.

\begin{figure}[h]
\centering
{\includegraphics{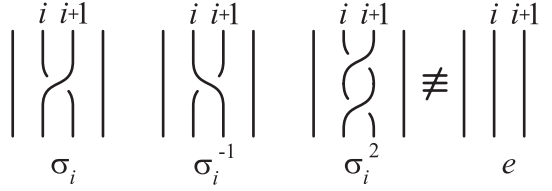}}
\caption{\label{fig:1} The geometrical presentation of the generator $\sigma_i$ of the full braid group for $R^2$ and its inverse $\sigma_i^{-1}$ (left); in 2D $\sigma_i^2\neq e$ (right)}
\end{figure}

It is well known \cite{wilczek,wu,sud}, that the source of exotic Laughlin correlations is of a 2D peculiar topology-type. This special topology of planar systems is linked with exceptionally rich structure of braid groups for 2D manifolds (like $R^2$, or compact manifolds like sphere or torus) in comparison to braid groups for higher dimensional spaces ($R^d$, $d>2$) \cite{birman}. The full braid group is defined as $\pi_1$ homotopy group of the $N$--indistinguishable--particle configuration space, i.e., the group of multi-particle trajectory classes, disjoint and topologically nonequivalent (trajectories from various classes cannot be continuously transformed one onto another one). The full braid groups are infinite for 2D case while are finite (and equal to the ordinary permutation group $S_N$) in higher dimensions of the manifold on which particles can be located \cite{birman,jac}. This property makes two dimensional systems  exceptional in geometry--topology sense. For matching the topological properties with quantum system properties, the quantization according to the Feynman path integral method is especially  useful \cite{wilczek,wu,lwitt}. Due to a fundamental ideas of path integral quantization in the case of not simply connected configuration spaces (indicated by nontrivial $\pi_1$ group), like for the multi-particle systems, additional phase factors---weights of nonequivalent (nonhomotopic) trajectory classes and summation over these classes must be included (a measure in the trajectory space is distributed over separated disjoint homotopy classes of $\pi_1$). As it was proved in Ref. [\onlinecite{lwitt}], these weight factors form a one-dimensional unitary representation (1DUR) of the full braid group. Different 1DURs of the full braid group give rise to distinct types of quantum particles corresponding to the same classical ones. In this manner one can get fermions and bosons corresponding to only possible 1DURs of $S_N$, $\sigma _i \rightarrow e^{i\pi}$ and $\sigma_i\rightarrow e^{i0}$, respectively, (the permutation group $S_N$ is the full braid group in 3D and in higher dimensions, $\sigma_i$, $i=1,...,N$ denote generators of $S_N$). For more rich braid group in 2D one encounters, however, the infinite number of possible so-called anyons (including bosons an fermions) related to 1DURs, $\sigma_i\rightarrow e^{i\Theta}$, $\Theta \in [0,2\pi)$ ($\sigma_i$ are here generators of the full braid group in 2D, cf. Fig. \ref{fig:1}) \cite{wilczek,wu,sud,birman}. 

CFs associated with Laughlin correlations require, however, the statistical  phase shift $p \pi$, with $p=3,5,7...$ for its various types, and the periodicity of 1DURs, $e^{ip\pi}=e^\pi=-1$, does not allow to distinguish CFs from ordinary fermions. It caused some misinterpretation---CF fermions were treated \cite{jain,hon} as ordinary fermions dressed somehow with flux-tubes in analogy to solid-state quasiparticles, which is, however, an incorrect picture.

In the present paper we revisit the topological approach to Hall systems and recover Laughlin correlations by employing properties of the underlying cyclotron braids \cite{jac1,jac2}, originally defined and without a phenomenological modeling of CFs. We will demonstrate that particles with statistical properties of CFs are not composites of fermions with flux-tubes or vortices, but are rightful 2D quantum particles characterized by 1DURs of cyclotron braid subgroups. We notice also that the original CFs construction with flux-tubes employing a heuristic assumption that the mean field of local fluxes can be greater than the external, to be justified in terms of cyclotron braid subgroups needs a special assumption on possible quantum dynamics. One can formulate, however, the  FQHE hierarchy in terms of IQHE in resultant field, avoiding the previously made assumption on possible $\pm$ sign of the effective field. The explanation of the mechanism for creation of the effective field for fractional fillings of LL in terms of cyclotron braid groups would be also helpful for identification of Chern-Simons field constructions \cite{cs}, which were widely spread in modeling of CFs and anyons within mathematical effective approach to Hall systems in fractional regime.

\section{Too-short for interchanges cyclotron trajectories in 2D Hall systems}

One-dimensional unitary representations (1DURs) of full braid group \cite{wu,birman,imbo}, i.e., of $\pi_1$ homotopy group of the configuration space for indistinguishable $N$ particles \cite{birman}, define weights for the path integral summation over trajectories \cite{,wilczek,wu,lwitt}. If trajectories fall into separated homotopy classes that are distinguished by non-equivalent closed loops (from $\pi_1$) attached to an open trajectory $\lambda_{a,b}$ (linking in the configuration space points, $a$ and $b$), then an additional summation over these classes with an appropriate unitary factor (the weight of the particular trajectory class) should be included \cite{wilczek,wu} in the path integral (for transition from the point $a$ at the time moment $t=t_1$ to the point $b$ at $t=t_2$):
\begin{equation}
I_{a,t_1\rightarrow b,
t_2}=\sum\limits_{l\in\pi_1} e^{i\alpha_l}\int
d\lambda_l e^{iS[\lambda^{l}_{(a,b)}]},
\end{equation}
where $\pi_1$ stands for the full braid group and index $l$ enumerates $\pi_1$ group elements, $\lambda^l$ indicates an open trajectory $\lambda$ with added  $l$th loop from $\pi_1$ (the full  braid group here). The factors $e^{i\alpha_l}$ form a 1DUR of the full braid group and distinct representations
correspond to distinct types of quantum particles \cite{wu,lwitt}. The closed loops from the full braid group describe exchanges of identical particles, thus, the full braid group 1DURs indicate the statistics of particles \cite{wilczek,wu,sud}.

Nevertheless, it is impossible to associate in this manner CFs with the 1DURs of the full braid group, because 1DURs are periodic with a period of $2\pi$, but CFs require the statistical phase shift of $p\pi$, $p=3,5...$. In order to solve this problem, we propose \cite{jac1} to associate CFs with appropriately constructed braid subgroups instead of the full braid group and in this way to distinguish CFs from ordinary fermions.

The full braid group contains all accessible closed multi-particle classical trajectories, i.e., braids (with initial and final orderings of particles that may differ by permutation, which is admitted for indistinguishable particles). One can, however, notice that inclusion of a magnetic field substantially changes trajectories---a classical cyclotron motion confines a  variety of accessible braids. When the separation of particles is greater than twice the cyclotron radius, which situation occurs at fractional lowest LL fillings, the exchanges of particles along single-loop cyclotron trajectories are {\it precluded}, because the cyclotron orbits are {\it too short}  for particle interchanges. Particles must, however, interchange in the braid picture for defining the statistics and in order to allow exchanges again, the cyclotron radius must somehow be {\it enhanced}. The natural way is to exclude inaccessible braids from the braid group. We will show  that remaining braids would be sufficient for particle exchanges realization.

One can argue that cyclotron radius enhancement could be achieved by either lowering the effective magnetic field or lowering the effective particle charge. These two possibilities lead to the two phenomenological concepts of CFs---with the lowered field in Jain's construction \cite{jain} and with the screened charge in Read's construction of vortices \cite{vor3}. Both these constructions seem to not matter with braid groups, but actually both of these effective phenomenological tricks correspond to the same, more basic and natural concept, of restricting the braid family by excluding inaccessible trajectories \cite{jac1,jac2}. We will demonstrate below that at sufficiently high magnetic fields in 2D charged $N$--particle systems, the {\it multi-looped braids} allow for the effective enlargement of cyclotron orbits, thus restoring particle exchanges in a natural way \cite{jac2}. These multi-looped braids form a subgroup of the full braid group and, in the presence of strong magnetic field, the summation in the Feynman propagator will be thus confined to the elements of this subgroup (its semigroup, for fixed magnetic field orientation, however, with the same 1DURs as of the subgroup).

\section{Cyclotron braid subgroups---restitution of particle interchanges}

\begin{figure}[h]
\centering
{\includegraphics{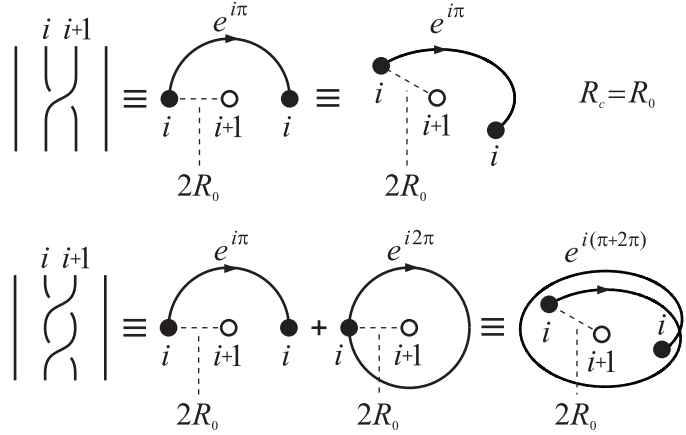}}
\caption{\label{fig3} The generator $\sigma_i$ of the full braid group and the corresponding relative trajectory of the $i$th and $(i+1)$th particles exchange (upper); the generator of the cyclotron braid subgroup, $b_{i}^{(p)}=\sigma_i^p$ (in the figure, $p=3$), corresponds to additional $\frac{p-1}{2}$ loops when the $i$th particle interchanges with the $(i+1)$th one  (lower) ($2R_0$ is the inter-particle separation, $R_c$ is the cyclotron radius, 3D added for better visualization)}
\end{figure}

Mentioned above multi-looped  braids form the {\it cyclotron braid subgroups} and are generated by the following generators:
\begin{equation}
\label{gen}
b_i^{(p)}=\sigma_i^p,
\;\;(p=3,5...),\;i=1,...,N-1,
\end{equation}
where each $p$ corresponds to a different type of the cyclotron subgroup
and $\sigma_i$ are the generators of the full braid group. The group element $b_i^{(p)}$ represents the interchanges of the $i$th and $(i+1)$th particles
with $\frac{p-1}{2}$ loops, which is clear by virtue of the definition of the
single interchange $\sigma_i$ (cf. Fig. \ref{fig3}, e.g., for $p=3$ one deals with elementary particle exchange braid with one additional loop). It is clear that $b_i^{(p)}$ generate a subgroup of the full braid group as they are expressed by the full braid group  generators $\sigma_i$.

The 1DURs of the full group confined to the cyclotron subgroup (they do not depend on $i$ as 1DURs of the full braid group do not depend on $i$ by virtue of the $\sigma_i$ generators property, $\sigma_i\sigma_{i+1}\sigma_i=\sigma_{i+1}\sigma_i \sigma_{i+1},\;\; 1\leq i\leq N-1$ \cite{birman,jac}) are 1DURs of the cyclotron subgroup:
\begin{equation}
\label{repr}
b_i^{(p)}\rightarrow e^{ip\alpha},\;i=1,...,N-1,
\end{equation}
where $p$ is  an odd integer and $\alpha \in (-\pi,\pi]$. We argue, that these 1DURs, enumerated by the {\it pairs} ($p$, $\alpha$), describe composite anyons (CFs, for $\alpha =\pi$). Thus in order to distinguish various types of composite particles one has to consider ($p$, $\alpha$) 1DURs of cyclotron braid subgroups.

In agreement with the general rules of quantization \cite{sud,imbo}, the $N$-particle wave function must transform according to the 1DUR of an appropriate element of the braid group, when the particles traverse, in classical terms, a closed loop in the configuration space corresponding to this particular braid element. In this way the wave function acquires an appropriate phase shift due to particle interchanges (i.e., due to exchanges of its variables according to the prescription given by braids in 2D configuration space). Using 1DURs as given by (\ref{repr}), the Aharonov-Bohm phase of Jain's fictitious fluxes is replaced by contribution of additional loops (each loop adds $2\pi$ to the total phase shift, if one considers 1DUR with $\alpha=\pi$ related to CFs, cf. Fig. \ref{fig3} (right)). Let us emphasize that the real particles do not traverse the braid trajectories, as quantum particles do not have any trajectories, but exchanges of coordinates of the $N$-particle wave function can be represented by braid group elements; in 2D a coordinate exchange do not resolve itself to permutation only, as it was in 3D, but must be performed according to an appropriate element of the braid group, being in 2D not the same as the permutation group \cite{wu,sud,imbo}. Hence, for the braid cyclotron subgroup generated by $b_i^{(p)}$, $i=1,...,N-1$, we obtain the statistical phase shifts $p\pi$ for CFs (i.e., for $\alpha=\pi$ in Eq. (\ref{repr})), as required by Laughlin correlations, without the need to model them with flux tubes or vortices.

\begin{figure}[h]
\centering
\scalebox{0.7}{\includegraphics{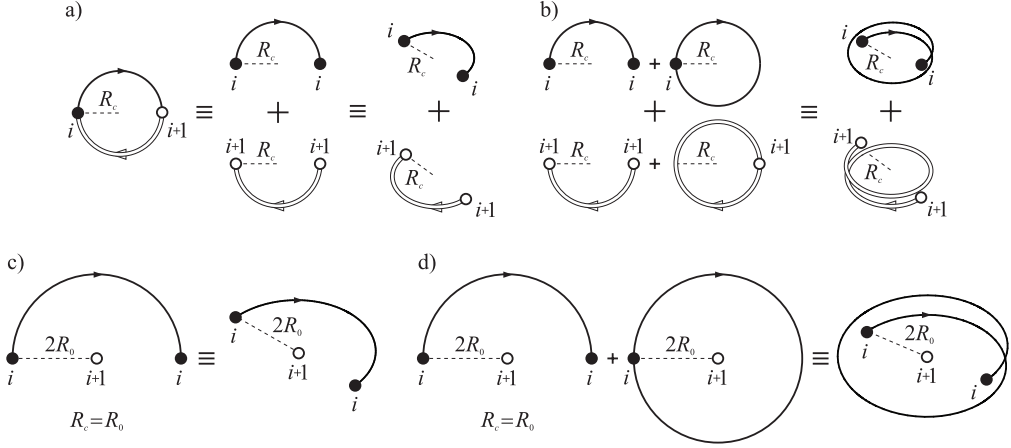}}
\caption{\label{fig4} Half of the individual particle cyclotron trajectories of the $i$th and $(i+1)$th particles (top) and the corresponding relative trajectories (bottom) for interchanges of the $i$th and $(i+1)$th 2D-particles under a strong magnetic field, for $\nu =1$ (left) and for $\nu =\frac{1}{3}$ (right), respectively ($R_c$---cyclotron radius, $2R_0$---particle separation, 
3D added for better visualization)}
\end{figure}

\begin{figure}[h]
\centering
\scalebox{0.7}{\includegraphics{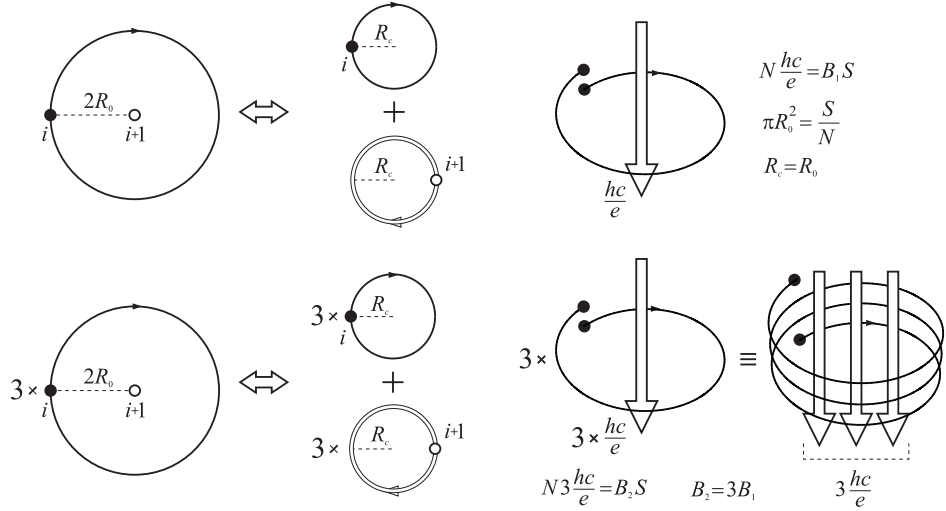}}
\caption{\label{fig5} Cyclotron trajectories of individual particles must be closed, therefore they correspond to 
{\it double} exchange braids,  for both,  simple exchanges (upper) and  exchanges with additional loops (lower), in the right part, quantization
of flux per particle, for $\nu=1$ and $\nu=\frac{1}{3}$, is indicated}
\end{figure}

Each additional loop of a relative trajectory for the particle pair interchange (as defined by the generators $b_i^{(p)}$) reproduces an additional loop in the individual cyclotron trajectories for both interchanging particles---cf. Fig. \ref{fig4}. The cyclotron trajectories are repeated in the relative trajectory (c,d) with twice the radius of the individual particle trajectories (a,b). In quantum language, with regards to classical multi-looped cyclotron trajectories, one can conclude only on the number, $\frac{BS}{N}/\frac{hc}{e}$, of flux quanta per single particle in the system, which for the filling $\frac{1}{p}$ is $p$ (for odd integer $p$), i.e., the same as the number of individual particle cyclotron loops (which equals to $p=2n+1$, where $n=1,2...$ indicates the number of additional braid-loops for particle interchange trajectories). From this observation it follows a simple rule: for $\nu =\frac{1}{p}$ ($p$ odd), each additional loop of a cyclotron braid corresponding to particle interchange, results in {\it two} additional flux quanta piercing the individual particle cyclotron trajectories. This rule follows immediately from the definition of the cyclotron trajectory, which must be a {\it closed} individual particle trajectory related to a {\it double} interchange of the particle pair (cf. Fig. \ref{fig5}). In this way, the cyclotron trajectories of both interchanging particles are closed, just like the closed relative trajectory for the double interchange (the braid trajectory is open as the trajectory of particle interchange only, and therefore the double interchange is needed to close this trajectory). If the interchange is simple, i.e., without any additional loops, the corresponding individual particle cyclotron trajectories are also simple, i.e., single-looped. Nevertheless, when the interchange of particles is multi-looped, as associated with the $p$-type cyclotron subgroup ($p>1$), the double interchange relative trajectory has $2 \frac{p-1}{2}+1=p$ closed loops, and the individual cyclotron trajectories are also multi-looped, with $p$ loops \cite{jac2,EPL}.

All these properties of multi-looped planar trajectories at strong magnetic field are linked with the fact that in 2D additional loops cannot enhance the total surface of the system. In this regard, it is important to emphasize the basic difference between the turns of a 3D winding (e.g., of a wire) and of multi-looped 2D cyclotron trajectories. 2D multi-looped trajectories do not enhance the surface of the system and therefore do not enhance total magnetic field flux $BS$ piercing the system, in opposition to 3D case. In 3D case, each turn of a winding adds a new portion of flux, just as a new turn adds a new surface, which is, however, impossible in 2D. Thus in 2D all loops must share the same total flux, which results in {\it diminishing} flux-portion per a single loop and, effectively, in longer cyclotron radius (allowing again particle interchanges).

The additional loops in 2D take away the flux-portions (equal to $p-1$ flux quanta just at $\nu=\frac{1}{p}$, $p$ odd) simultaneously diminishing the effective field; this gives an explanation for Jain's auxiliary fluxes screening the external field $B$. Thus, it is clear that CFs are actually not compositions of particles with flux-tubes, but are rightful particles in 2D corresponding to 1DURs of the cyclotron subgroups instead of the full braid group, which is unavoidably forced by too short ordinary single-looped cyclotron trajectories. The original name 'composite fermions' can be, however, still used. Moreover, one can use a similar name, 'composite anyons', for particles associated with fractional 1DURs (i.e., with fractional $\alpha$) of the cyclotron subgroup instead of the full braid group, the latter linked rather with ordinary anyons (without magnetic field).

\section{The mapping of FQHE onto IQHE}

Let us emphasize that the agreement between loop number and flux quanta number per particle (allowing, in fact, for a formulation of Jain's model of CFs) is restricted only to fillings $\frac{1}{p}$, $p$ odd. Out of these fillings, the number of flux quanta per particle cannot be equal to number of loops as it is not an integer (while the number of loops is always integer). For fields out of $\nu=\frac{1}{p}$, on a single loop it falls non-integer number of flux quanta (it may happen, because cyclotron loops are classical braid-type objects, not quasiclassical trajectories with flux quantization requirements). 
In other words, all loops together must take away the total flux of the external field. Differently it is assumed in CF construction with concept of rigid flux quanta attached to particles even out of $\nu=\frac{1}{p}$  filling fraction. In particular, it is assumed by Jain that the resultant effective field of the fictitious flux quanta and of the external magnetic field would be even negative (oriented oppositely to the external filed) as e.g., for $\nu >\frac{1}{2}$ (for $p=3$) \cite{jain,hon}. The concept of Jain's resultant field leads to  the FQHE hierarchy obtained via mapping of FQHE onto IQHE, $\nu=\frac{n}{(p-1)n\pm 1}$ \cite{jain}.

One can, however, consider the mapping of FQHE onto IQHE within cyclotron braid approach, assuming uniform distribution of the external field flux over all trajectory loops. From point of view of multi-looped cyclotron braids, in the case of $\nu\neq \frac{1}{p}$ ($p$ odd), on each loop it falls a fraction of a flux quantum and if it coincides with the same fraction as per single particle for completely filled several Landau levels (with single cyclotron loops) the mapping of IQHE onto FQHE holds, resulting in filling hierarchy. One can compare the flux-fractions per single loop, for fractional and integer LLs fillings:
\begin{equation}
\begin{array}{llll}
FQHE: & \nu=\frac{N}{N_0},& N_0=\frac{BS}{hc/e},&
\Phi_F=\frac{BS}{Np}=\frac{hc}{e\nu p},\\
IQHE (n-th\;\; LL):& n=\frac{N_1}{N_0},&
N_0=\frac{B_1S}{hc/e},& \Phi_I=\frac{B_1
S}{N_1}=\frac{hc}{en},\\
\end{array}
\end{equation}
and, in the case when the flux per single loop in FQHE, $\Phi_F =\frac{BS}{Np}=\frac{hc}{e\nu p}$ is equal to the flux per single particle  in IQHE (thus, per single loop, as for IQHE cyclotron trajectories are single-looped), $\Phi_I=\frac{B_1 S}{N_1} =\frac{hc}{en}$, the mapping of FQHE onto IQHE holds, and it happens when $\nu=\frac{n}{p}$,  where $n=1,2,3,4...$, $p=1,3,5...$. This reproduces FQHE hierarchy (due to the mapping onto IQHE) avoiding problems with  sign minus in the former formula $\nu=\frac{n}{(p-1)n\pm 1}$ \cite{jain}.

For filling rates out of $\frac{1}{p}$, $p$--odd, one deals with still integer number of additional loops per particle but not with integer number of flux quanta. Any flux-tubes attached to composite fermions do not exist, they are only a convenient model for additional loops in classical cyclotron braid picture (allowing interchanges of particles when single-looped cyclotron trajectories are too short) and exceptionally for $\nu=\frac{1}{p}$ ($p$--odd) they would be imagined as of $p-1$ flux quanta attached to particles and oppositely oriented  to external field, but out of these fillings, not.

Nevertheless, in order to rescue in terms of multi-looped trajectories the assumption of Jain, that even out of $\nu=\frac{1}{p}$, rigid flux quanta are associated to particles, one can consider the situation when all but last one of particular loops from multi-loop structure embrace the integer number of flux quanta and only the last loop takes away a fractional rest of the total flux. This needs, however, the conjecture to be formulated, that the braid multi-looped structure is repeated by semiclassical quantum dynamics of wave-packets, thus rigorously satisfying the flux quantization rule. Formation of such wave-packets traversing closed cyclotron loops (in order to define the magnetic field flux trough their orbits) is highly sophisticated, as single particle quantum dynamics might not manifest cylindrical symmetry (depending of a gauge choice) \cite{landau1972}, and operators corresponding to classical position of cyclotron orbit center ($x$ and $y$) do not commute \cite{eliu}. Nevertheless, the quantum evolution of the position ($x$, $y$) operators and the corresponding momentum operators, in the Heisenberg picture, is cyclic in time in the presence of the $z$-axis oriented magnetic field. This leads to a periodic (with cyclotron frequency) evolution in the plane perpendicular to the magnetic field of any wave-packet of single-particle stationary states  \cite{eliu}. This property is conjectured to be conserved also in a strongly interacting many particle system collective states.

In the case when  the rest of the total flux per the last loop is fractional, the flux quantization rule is not fulfilled with regard to this loop. Nevertheless, if the rest of the flux per the last loop  is equal to $n$-th part of the flux quantum (with sign minus or plus, which  happens out of $\frac{1}{p}$), one could expect some special organization of quantum dynamics in  the form of the orbital movement of semiclassical wave-packets. These wave-packets would represent  particles and  must satisfy the rigid quantization of  the external magnetic field flux passing trough the surfaces of closed wave-packet semiclassical orbits. The rest flux per the last orbit is apparently not an integer quantum and even would be positive or negative, depending of the orientation of the resultant Jain's field with respect to the external field. In the case of this rest flux equals to $\pm \frac{hc}{ne}$, in  order to fulfill the flux quantization rule, one could, however, imagine an organizing  of the collective $n$ particle closed trajectory embracing the flux quantum in analogy to quantum dynamics for completely filled $n$ Landau levels. Here one can invoke a pictorial interpretation of the cyclotron trajectory fitted in length to $n$ de Broglie waves. In this way all loops would embrace the rigid quanta of magnetic field flux. 

The negative sign of the rest flux, $\pm \frac{hc}{ne}$, passing through the last loop of the individual particle (i.e., before collectivization of $n$ particles in order to satisfy the flux quantization rule) would  mean here the opposite direction (with respect to remaining loops, contrary to the direction induced by the external field) movement, resulting in the something like eight-shape multi-looped cyclotron structure of eventual wave-packet trajectory. Such strange picture of possible arrangement of quasiclassical movement for this particular LL fillings (given by Jain's FQHE hierarchy $\nu=\frac{n}{n(p-1)-1}$) satisfies, however, the requirements of  the flux quantization and explains the heuristic assumption of Jain's composite fermion construction with rigid flux quanta attached to particles. 

In order to verify  the above formulated conjecture interesting would be a measurement of cyclotron focusing of tiny beam of 2D ballistic carriers passing trough a nanometer-scale slot \cite{ogn}. This beam bent by the magnetic field to the left or to the right with respect to the source slot, for two mentioned above directions of cyclotron movement in the last loop, could be observed via resonant signal in nearby target slots in the case of commensuration of the cyclotron radius and the separation of slots  \cite{ogn,hon}. An asymmetry should be observed when passing $\nu=\frac{1}{2}$ LL filling by lowering or rising  the magnitude of the external field. This would be helpful in experimental confirmation of one of two above presented possibilities with either  uniform distribution of the external field flux over all loops, as for classical ones, or with nonuniform distribution forced by the quasiclassical rule of flux quantization, if actually the classical multi-looped cyclotron trajectories are repeated by wave-packets. 

Supporting to this idea are also recent experimental results regarded observations of FQHE in graphene \cite{fqhe1,fqhe2}. These observations are enabled by the special preparation of so-called suspended graphene sheets without contact with substrate matter and thus with extremely high value of carrier mobility (above 200 000 cm$^2$V$^{-1}$s$^{-1}$), which will be described in details in section \ref{graphen} of the present paper. Such high level of mobility apparently  favors  multi-loop dynamics of quasiclasical wave-packet and seems to allow for FQHE observations in relatively high temperatures (even up to 20 K; the another favorable circumstance is high value of cyclotron energy in graphene, by two order higher than in semiconductor heterostructures, enabling observation of IQHE even at room temperatures \cite{gr2}). In Ref. [\onlinecite{fqhe2}] the observation of FQHE at $\nu = \frac{1}{3}$ is reported at the field 12-14 T for the carrier concentration 10$^{11}$ 1/cm$^{2}$, while in Ref. [\onlinecite{fqhe1}] at much lower field 2-12 T for the concentration 10$^{10}$ 1/cm$^{2}$. This agrees with the concept of too short cyclotron radius, since for lower concentration, when separation of particles is bigger, it occurs at lower fields (in graphene the concentration of carriers can be  controlled by lateral gate voltage determining a position of the Fermi level), despite that  for lower concentration interaction of carriers is weaker due to particle density dilution. These tendencies of FQHE seem to support  cyclotronic scenario for FQHE and not the overwhelming role of the interaction. The next argument  follows also from the reported observation of FQHE concurrence with insulating state  in suspended graphene \cite{fqhe1} (interpreted  as Wigner crystalization \cite{wigner}; in the case of semiconductor heterostructures it was shown \cite{wig10} that for $\nu<\frac{1}{9}$ the Laughlin state is energetically less favorable than the state with localized  electrons---this agrees also with              
multi-looped cyclotron dynamic picture since kinetic energy grows  proportionally to loop number, more rapidly than for single-looped trajectories). In graphene the insulating state manifests itself at very low concentrations when the energetically inconvenient high value of loops would be necessary to match distantly separated particles even at moderate high magnetic fields \cite{fqhe2} (interesting is that annealing of the graphene sample enhances stability of FQHE versus insulating state, probably because reducing  defect rate, which in turn  enhances the  mobility of carriers). The above experimental evidences would suggest that the multi-looped braid picture is somehow repeated in true dynamics with energy consequences and with high mobility of carriers requirements.    

The other problem raised by cyclotron braid approach consists in the fact that CF are not ordinary fermions dressed with interaction, but are separated 2D quantum particles, and they cannot be mixed with ordinary fermions (similarly as bosons cannot be mixed with fermions), especially within numerical variational interaction minimizations or diagonalizations. Even though both fermions and CFs correspond to antisymmetric wave functions, not all antisymmetric wave functions describe CFs and the domain for minimization can comprise only these antisymmetric functions which transform according to appropriate 1DUR of the cyclotron subgroup (it is a subspace of the Hilbert space of antisymmetric functions). The minimizations done on the whole domain of antisymmetric functions would lead thus to improper results.
\section{The influence of the Coulomb interaction}
The Coulomb interaction play a central role for Laughlin correlations \cite{haldane,prange,laughlin1}, but in 2D systems upon the quantized magnetic field, the interaction of charges cannot be accounted for in a manner of standard dressing of particles with interaction as it was typical for quasiparticles in solids, because in 2D Hall regime this interaction does not have a continuous spectrum with respect to particle separation expressed in relative angular momentum terms \cite{haldane,prange}. The interaction can be operationally included within the Chern-Simons (Ch-S) field theory \cite{cs,lopez}, formulating an effective description of the local gauge field attached to particles, which, in the area of Hall systems, suits to particles with vortices, such as anyons and CFs \cite{hon}. It has been demonstrated \cite{prange,mo} that the short-range part of the Coulomb interaction stabilizes CFs against the action of the Ch-S field (its antihermitian term \cite{mo,vor16}), which mixes states with distinct angular momenta within LL \cite{mo}, in disagreement with the CF model in the Ch-S field approach \cite{hon,mo}. The Coulomb interaction removes the degeneracy of these states and results in energy gaps which stabilize the CF picture, especially effectively for the lowest LL. For higher LLs, the CFs are not as useful due to possible mixing between the LLs induced by the interaction \cite{id}. The short-range part of the Coulomb interaction also stabilizes the CFs in cyclotron braid terms \cite{jac1}, similarly to how it removes the instability caused by the Ch-S field for angular momentum orbits in LL \cite{mo}. Indeed, if the short-range part of the Coulomb repulsion was reduced, the separation of particles would not be rigidly kept (adjusted to a density only in average) and then other cyclotron trajectories, in addition to those for a fixed particle separation (multi-loop at $\nu=\frac{1}{p}$), would be admitted, which would violate the cyclotron subgroup construction.

\section{Read's CFs and Hall metal state in cyclotron braid terms}

For Read's CFs \cite{vor3,vora1}, Laughlin correlations are modeled by collective vortices that are attached to the particles. A vortex with its center at $z$ is defined as \cite{vor3},
\begin{equation}
\label{vor1}
V(z)=\prod_{j=1}^{N} (z_j-z)^q,
\end{equation}
where $q$ is the vorticity. For odd $q$, it is linked to the Jastrow factor of the LF \cite{laughlin2}, $\prod_{i>j}^{N} (z_j-z_i)^q$, (resulting from Eq. (\ref{vor1}) by the replacement of $z$ with $z_i$ and the addition of $i$ ($i>j$) to the product domain, i.e., by the binding of vortices to electrons). In particular, for $q=1$ one arrives at the Vandermonde determinant,  $\prod_{i>j}^{N} (z_j-z_i) $ (being the polynomial part of the Slater function of $N$ noninteracting 2D fermions at magnetic field corresponding to $\nu=1$, i.e., to the case of the complete filled lowest LL), associated with the ordinary single-looped cyclotron motion of $N$ fermions on the plane at $\nu=1$. Because the vortices are fragments of the LF, they contain more information than just the statistical winding phase shift (the latter expressed by the factor, $\prod_{i,j}(z_i-z_j)^q/|z_i-z_j|^q$). 1DURs of the cyclotron braid subgroups define the statistical phase winding, but not the shape of the wave function. The wave function shape is determined via the energy competition between various wave functions with the same statistical symmetry. Thus, vortices contain information beyond just the statistical phase shift, they also include the specific radial dependence of multi-fold zeros pinned to particles through the Jastrow polynomial. The vortex is a collective fluid-like concept that does not meet the single-particle picture. The vorticity $q$ is selected, however, in accordance with the {\it known in advance} LF, thus, similarly as CF flux tubes, it requires a motivation within the cyclotron structure.

The properties of vortices can be listed as follows \cite{vor3}:
\begin{itemize}
\item{ when traversing with an arbitrary particle $z_j$ a closed loop around the vortex center, then the gain in phase is equal to $2\pi q$;}
\item{the vortex induces a depletion of the local charge density, which results in a locally positive charge (due to background jellium) that screens the charge of the electron associated to the vortex center; this positive charge is $-q\nu e$ (for $\nu =1/q$ it gives $-e$, which would completely screen the electron charge);}
\item{exchange of vortices results in a phase shift of $q^2\nu \pi$, (due to the charge deficit of the vortex), which for $\nu=\frac{1}{q}$ gives $ q\pi$; the $q$-fold vortex, together with the bound electron (which contributes a charge $e$ to the complex and produces a statistics phase shift of $\pi$), form a complex that behaves like a composite boson with zero effective charge for odd $q$ and like a composite fermion  for even $q$.}
\end{itemize} 
The bosons can condense and, in this manner, reproduce exactly the LF for odd $q$ \cite{vor16}, while, for even $q$, one deals with the Fermi sea in a zero net field, as in both cases the effective charge of the complexes is zero; the latter case corresponds to the Hall metal state \cite{halperin,vor7,vor11}.

The second property of listed above, explains why the model with vortices works. The reduced effective charge of the electron--vortex complex, results in an increase of the cyclotron radius, which is necessary for particle exchanges at fractional fillings.

The specific character of the concept of vortices is clearly visible, in particular, for $\nu=1$. Vortices of the form (\ref{vor1}) with vorticity $q=1$, attached to electrons in the system, result in the Vandermonde factor (being the Jastrow factor with exponential $q=1$). In this case, the corresponding Laughlin state is thus given by the Slater function of $N$ {\it noninteracting} fermions, what, however, can also be effectively described by the Bose Einstein condensate of bosons 
defined as fermions with vortices (\ref{transf}) with $q=1$ (all the action of the magnetic field on ordinary fermions is replaced with this Bose condensation). The Coulomb interaction do not contribute in this particular case, since at $\nu=1$ the Haldane pseudopotential \cite{haldane,prange} (i.e., the short range part of the   Coulomb interaction, being essential in the selection of the Laughlin state form) is zero (as $q-2<0$, for $q=1$), and thus the Slater function of {\it noninteracting} particles is suitable as the eigen-state of the interacting system at $\nu=1$.

A phenomenological modifications of vortices, like a shift of the centre of the vortex from the position of an associated electron, may result in effective attraction of vortex-composite fermions, leading to their pairing at e.g., $\nu =5/2$ \cite{vora1,vor17}. This corresponds, in fact, to a modification of the Laughlin function and leads to a new wave function, in this case, $N$ particle BCS-like function in the form of Pfaffian, as was described in Refs [\onlinecite{vora1,vor17}]. 
Note that the wave function with the Pfaffian factor is still of the same statistical symmetry as that for the particular sort of braid-composite fermions (defined by 1DUR of the corresponding cyclotron subgroup).

All properties of vortices or flux-tubes  in CF constructions can be grasped together by a formal local gauge transformation \cite{vor16} of the original fermion particles (defined by the fermion field operator $\Psi({\bf x})$) to composite particles represented by fields (annihilation and creation):
$
\label{transf}
\Phi({\bf x})=e^{-J({\bf x})}\Psi({\bf x}),\;\;\Theta({\bf
x})=\Psi^+({\bf x})e^{J({\bf x})}$, where:
$J ({\bf x}) =q\int d^2x'\rho({\bf
x}')log(z-z')-\frac{|z|^2}{4l^2}$, and $e^{-J}
$ 
corresponds to a nonunitary, in general, transformation that describes the attachment of Read's vortices (or Jain's flux-tubes) to the bare fermions, $ \Psi({\bf x})$ and $ \Psi^+({\bf x})$ (for the original fermion annihilation and creation fields, respectively). When restricting $J ({\bf x})$ to only its imaginary part (i.e., to the imaginary part of $log$), one arrives at the hermitian Ch-S field corresponding to the dressing of fermions with local flux-tubes   \cite{vor18}. The field operators $ \Phi({\bf x})$ and $ \Theta({\bf x})$, $\Phi^+({\bf x}) =\Theta({\bf x})e^{J({\bf x})+J^+({\bf x})}$, though are not mutually conjugated (they are perfectly conjugated for the hermitian Ch-S field),    describe composite bosons (for odd $q$) and composite fermions (for even $q$) within the mean field approach \cite{vor16} (remarkably, the real part of $J$ vanishes in the mean field, as the real part of $log$ is canceled by the Gaussian, while the hermitian Ch-S field is canceled by the external magnetic field). From the relation $e^{q\sum_{j}log(z-z_j)}=\prod_{j}^{N}(z-z_j)^q$ (for the density operator $\rho({\bf x})=\Psi^+({\bf x})\Psi({\bf x})\Longrightarrow \sum_{j=1}^{N}\delta(z-z_j)$), which coincides with the definition of Read's vortex, one can expect that the above local gauge transformation reproduces all properties of vortices. This gauge transformation allows for the interpretation of the Laughing state as a Bose-Einstein condensate of composite bosons, at $\nu = \frac{1}{q}$, $q$---odd \cite{vor3,vor16}, and as a compressible fermion sea, at $q$---even \cite{vor7,vor11} (the latter is unstable against BCS-like pairing) \cite{vora1,vor17}. Assuming that the CFs are defined by the 1DURs of the cyclotron subgroup, the hermitian term of this gauge transformation should be omitted, because it defines CFs when starting from ordinary fermions, which are already taken into account in terms of cyclotron braids.

Let us finally comment on the $\nu=\frac{1}{2}$ state (Hall metal) from the point of view of the braid approach.  Within Jain's model, two flux-tubes attached to composite fermions completely cancel an external magnetic field in the mean field approximation (in other words, the hermitian Ch-S field associated with Jain's model cancels, in mean field,  the external magnetic field), and this results in a Fermi sea, called the Hall metal state \cite{halperin}. Within Read's approach to composite particles at $\nu=\frac{1}{2}$, the complete cancellation of charge takes place due to the charge density depletion of the vortex with $q=2$.  Mutual interchange of 2-fold vortices produces $q^2\nu \pi=2\pi$ phase shift and including additional $\pi$ due to electrons, the complexes of 2-fold vortices with electrons behave like fermions (without charge)---thus form a Fermi sea (Hall metal). The instability of the Fermi system, results next in a paired state expressed by the Pfaffian factor, restoring incompressibility due to the pairing-gap (BCS-like paired state at $\nu = 5/2$ \cite{vor17,vor9}, also considered for $\nu = 1/2$ and $1/4$ \cite{kr,kr1}). As Pfaffian \cite{vora1} contributes with $-\pi$ to the phase shift due to particle interchanges, the total phase shift of the wave function with the Jastrow polynomial $\prod_{i>j}(z_i-z_j)^2$ \cite{vora1,vor17} is $\pi$. This phase is given by the 1DUR of the cyclotron braid group (with $p=3$, as such a cyclotron braid subgroup corresponds to the range $\nu \in [1/3,1)$) assigned by $p\alpha=3\frac{1}{3}\pi=\pi$, i.e., $\alpha=\frac{1}{3}\pi$.  The representation  ($p=3$, $\alpha=\frac{1}{3}\pi$) induces the fermion statistics phase shift of the many-particle wave function for $\nu =1/2$,  and in terms of braid-composite fermions, it corresponds to a net composite electron Fermi sea (since two loops take away the total external flux),  in consistence with the  local gauge transformation with $q=2$, thus reproducing fermions (starting from ordinary fermions) \cite{vor3,vor16}.

\section{Experimental evidences in graphen for 
carrier mobility  role for  triggering  of FQHE and related its multilooped structure}
\label{graphen}

Recent experimental investigations of the FQHE in graphene \cite{fqhe2,fqhe1} have shed  new light on this correlated state and seem to go beyond explanative ability of CF treatment that concentrates solely on the interaction.  If one imagines CFs to be analogous to solid-state Landau quasiparticles dressed with the interaction, i.e., presuming that the  local flux tubes are a result of the interaction \cite{hon,jain2007}, one would lose  important topological effects and  encounter problems with new observations indicating that the carrier mobility (and not the interaction) plays a triggering role for the FQHE  in suspended graphene samples.

\subsection{IQHE and FQHE in graphene---description of experimental results}

A single-atom-thick layer of  graphene has a hexagonal 2D structure
 with two carbon atoms per unit cell.
This results in a double triangular lattice of carbon atoms arranged in a honeycomb pattern. The $p$ orbitals perpendicular to the plane hybridize to type $\pi$ of the band structure well described in the approximation of a strong coupling,   $E_{\pm}({\bm k}) = \pm t \sqrt{3+f({\bm k})} - t'f({\bm k})$, where $f({\bm k})=2\cos(\sqrt{3}k_{y}a)+ 4\cos(\frac{\sqrt{3}}{2}k_{y}a)\cos(\frac{3}{2}k_{x}a)$, $a\simeq 0.142$ nm is distance between carbon atoms, $t=2.7$ eV---hopping energy to the nearest neighbors (between sublattices), $t'=0.2t$---hopping energy to next-nearest neighbors (inside the sublattices). 
The valence band and the conduction band  meet in points called $K$ and $K'$ at the border of a hexagonal Brillouin zone \cite{gr1,gr2} in compliance with the above relation for $t'=0$. Both
bands met in these points (non-gap semiconductor) have  locally a conical shape, which means that the relation between energy and momentum (distance from points of contact) is linear and the appropriate band Hamiltionian  is formally equivalent to the  relativistic fermions with zero rest mass ($E=\pm\sqrt{m_0^2v_F^4+p^2v_F^2}$, with $m_0=0$), described by Dirac equation with the velocity of light replaced by the Fermi velocity, $v_F\simeq c/300$ \cite{gr2,yang}. Therefore, the dynamics equation looks as follows,
$
-i v_F \vec{\sigma} \cdot \nabla \Psi({\bm r})=E\Psi({\bm r}),
$
where the Pauli matrix vector corresponds to the pseudospin structure related to two sublattices \cite{gr2,gr3} (wave functions are spinors in this structure). The zero mass of the Dirac fermions leads to numerous consequences and electron anomalies in the properties of graphene \cite{gr2,gr3,gr4,gr5}. For Dirac particles with zero rest mass, momentum uncertainty also leads to energy uncertainty (contrary to non-relativistic case), which results in the time evolution mixing together particle states with hole (anti-particle)  states for relativistic type dynamics. For zero-mass Dirac electrons the scaling of cyclotron energy is different as well ($\sim B^{1/2} $, and not $\sim B$, as in the case of non-relativistic particles). The value of this energy is also different, and larger by far (two orders of magnitude larger than the one corresponding in classical materials, i.e., it is [due to zero mass in Dirac point] as much as about 1000 K, for 10 T field), which allows to observe the IQHE in graphene even at room temperatures \cite{gr4,gr5}. There is, however, an anomalous IQHE observed here (for $\nu=\pm 4(n+1/2)$, or for $\pm 2, \pm 6, \pm 10, \ldots$ and at zero Landau level in the Dirac point, i.e., for zero energy; $\pm$ corresponds to particles and holes, respectively,  4 results from pseudospin/valley degeneration, 1/2 is associated with Berry's phase for pseudospin) \cite{gr2,yang,gr3,gr4,gr5,clure}. The  Klein paradox, referring to ideal tunneling of Dirac particles by rectangular potential barriers leads to extensive mobility of charge carriers in graphene, which is experimentally observed even near Dirac point (Fermi level at the border between electrons and holes). In this point, the density of charges is zero (and the zero Landau level is located here, employing both bands) \cite{gr2,gr4,gr5,yang}.
\begin{figure}[h]
\centering
\scalebox{0.6}{\includegraphics{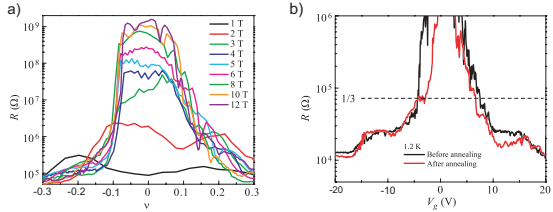}}
\caption{\label{grafenrys4}a) The emergence of an insulator state accompanying the increase in the strength of a magnetic field around the Dirac point, b) competition between the FQHE and the insulator state for the filling $-1/3$: annealing removes pollution which enhances the mobility and provides conditions for the emergence of a {\it plateau} for the FQHE, source Ref. [\onlinecite{fqhe2}]}
\end{figure}

 Despite using very strong magnetic fields (up to 45 T), FQHE was not detected in graphene samples deposited on a substrate of $SiO_2$ \cite{dur1}. In Ref. [\onlinecite{dur1}] it was  noted, however, the emergence of additional plateaus  of IQHE for the fillings $\nu = 0,\pm 1, \pm 4$, indicating the elimination of spin-pseudospin degeneration (related to sublattices), as a result of increasing mass of Dirac fermions \cite{dur1}. Only after mastering the novel technique of the so-called suspended ultrasmall graphene scrapings with extreme purity and high mobility of carriers (beyond 200000 cm$^2$V$^{-1}$s$^{-1}$; note that high mobility is necessary also to observe the  FQHE  in the case of semiconductor 2D hetero-structures), was it possible to observe the FQHE in graphene at fillings $\nu=1/3$ and $-1/3$ (the latter for holes, with opposite polarization of  the gate voltage, which determines Fermi level position, either in the conduction band, or in the valence band) \cite{fqhe1,fqhe2}. Both papers report the observation of the FQHE in graphene for strong magnetic fields. In the paper [\onlinecite{fqhe1}], in a field of $12-14$ T , for electron concentration of $10^{11}$/cm$^2$ (the mobility of 250000 cm$^2$V$^{-1}$s$^{-1}$) and in the paper [\onlinecite{fqhe2}], in a field of $2-12$ T, but for a concentration level smaller by one order  of magnitude ($10^{10}$cm$^{-2}$ and the mobility of 200000 cm$^2$V$^{-1}$s$^{-1}$). 

The FQHE in suspended graphene has been  observed at the temperatures around 10 K \cite{fqhe3}, and even higher (up to 20 K) \cite{fqhe6}. Authors of the paper [\onlinecite{fqhe6}] have argued that the critical  temperature elevation is related to the stronger electric interaction caused by the  lack of a dielectric substrate (with a relatively high dielectric constant in case of semiconductors, $\sim 10$) in the case of suspended samples. However, aspects that are likely  more important are the high mobility value (with suppressed acoustic phonon interaction in ideal 2D system, in comparison to 3D case) and,  on the other hand, the very high cyclotron energy in graphene (i.e., the large energy gap between the incompressible states). 

In the papers [\onlinecite{gr5,clure}] the competition between the FQHE state with the insulator state near the Dirac point has also been demonstrated, corresponding to a rapidly decreasing carrier concentration (and thus reducing interaction role at larger separation of carriers)---Fig.\ref{grafenrys4}.
 The most intriguing observation is that one \cite{fqhe2} demonstrating an influence of  annealing---in Fig.\ref{grafenrys4} b it is shown that FQHE occurs in the same sample originally insulating, upon same conditions, but after the annealing process enhancing mobility of carriers due to impurities reduction. This effect directly demonstrates the triggering role of carriers mobility in FQHE state arrangement.

\subsection{Quasiclassical quantization of the magnetic field flux for composite fermions}
\label{sec:2}

One can summarise topological main points related to FQHE: 
1) For fractional fillings of the LLL classical cyclotron orbits are too short for particle exchanges (as in the 2D case, long range helical motion is impossible). 
2) The exchanges are necessary to create a collective state such as the FQHE---thus the cyclotron radius must be enhanced (in the model of Jain's CFs \cite{jain} this enhancement is achieved by the artificial addition of flux tubes  directed oppositely to the external field; in the model of Read's vortices \cite{vor3}, reduction of the cyclotron radius is attained by the depletion of the local charge density---both this tricks are, however, effective and not supported in topological terms).
3) The topologically justified way to enhance cyclotron radius is with the use of multilooped cyclotron trajectories related to multilooped braids that describe elementary particle exchanges in terms of braid groups (the resulting cyclotron braid subgroup is generated by $\sigma_i^p$, $i=1,...,N$, $\sigma_i$ are generators of the full braid group \cite{jac1,jac2})---in 2D all loops of multilooped trajectory must share together the same total external magnetic field flux, in contrast to  the 3D case, and this is a reason of enhancement of all loops dimensions effectively fitting exactly to particle separation at LLL fillings $1/p$, ($p$-odd). 
4) In accordance with the rules of path integration for non-simply-connected configuration spaces \cite{lwitt},  one dimensional unitary representations (1DURs) of a corresponding braid group define the statistics of the system---in the case of multilooped braids, naturally assembled into the cyclotron subgroup \cite{jac2,EPL}, one arrives in this way at the statistical properties required by the Laughlin correlations (these 1DURs are $\sigma_i^p\rightarrow e^{ip\alpha},\;\alpha \in [0,2\pi)$, CFs correspond to $\alpha=\pi$).
5) The interaction is important for properly determining the cyclotron braid structure because its short range part  prevents the particles from approaching one another closer than the distance given by the density.

This completes a topological derivation of  FQHE statistics  without need to model it by  fictitious auxiliary elements like flux tubes, and proves that CFs are not quasiparticles dressed with interaction or complexes with local fluxes, but are rightful quantum 2D particles assigned with Laughlin statistics determined  by 1DURs of the appropriate cyclotron braid subgroup.

Nevertheless, since the model of CFs with attached rigid flux quanta  works so well (as evidenced by the exact diagonalizations) \cite{jain2007,hon}, the multilopped classical braid structure must be repeated by quasiclassical wave-packet trajectories (and then with quantized fluxes, which was not, however, a rule for classical  trajectories). Note that the flux quantization is the quasiclassical property as it needs a trajectory definition (the carries mobility also has the similar quasiclassical background). 

The wave packets corresponding to the  quasiclassical dynamics are  related to the collective character of a multiparticle system. For LLL of noninteracting system the group velocity of any packet is zero due to  degeneration of states. Interaction removes, however, this degeneracy and provides packet dynamics.  The collective movement minimizes kinetic energy, whereas the interaction favors localization (localization causes related increase in kinetic energy). Therefore, the collective dynamics  prefers the quasi-classical movement of packets along periodic closed trajectories (as a rule at the magnetic field presence \cite{eliu}), which then must, however, embrace  quantized external magnetic field fluxes. This is a role for collectivization in the energy preference of wave packets traversing closed trajectories in correspondence with the classical cyclotron description, including the multilooped braid picture \cite{jac2}.

This description appears to  be in accordance with the FQHE observations in graphene (described above), which are found at a low carrier density, and therefore accompany their dilution and the resulting reduction of interaction. Thus, the interaction is not the main factor initiating the FQHE, as was previously expected in view of the standard model of composite fermions, if one treated the dressing of fermions with localized flux tubes as a result of just the interaction itself \cite{hon,jain2007}.
Carrier mobility refers to semiclassical wave packet dynamics in terms of the drift velocity in an electric field and the classical Hall effect and reflects various channels of scattering phenomena beyond the simple model that includes only Coulomb interaction to free particles in the magnetic field. But topology arguments in the 2D case strongly prefer high mobility which is  required for real wave packets   to traverse multilooped trajectories as mobility is proportional to mean free path of carriers. Semiclassical wave packets even at the presence of the interaction and the scattering,  manifest periodic dynamics \cite{eliu} and in the case of the multilooped trajectory structure with the  enhanced radii the higher mobility is required, as has been experimentally demonstrated.

From the  cyclotron group pont of view, experimental results on FQHE in graphene \cite{fqhe1,fqhe2,fqhe3,fqhe6} seem to be compliant with the expectations of the braid description. In the case of suspended graphene,  controlling lateral gate voltage (within the range up to 10 V \cite{fqhe1}) allows regulating the density of carriers at a constant magnetic field. One should therefore expect  that at relatively  small densities of carriers (electrons, or symmetrical holes at reverse voltage polarization), the  cyclotron orbits will be too short to prevent braid exchanges of particles at a sufficiently strong magnetic field---although weaker for smaller concentrations---and experimental observations have supported  exactly  this prediction \cite{fqhe1,fqhe2}. For low concentration, while closing on the Dirac point, one may expect that too strong fields would exceed the stability threshold of the FQHE state in competition with the Wigner crystal (taking into account a specific character of this competition in the case of massless Dirac fermions in comparison to traditional semiconductor 2D structures \cite{wigno}) and that corresponds to the emergence of the insulating state near the Dirac point in a strong magnetic field. In the case of the hexagonal structure of graphene, electron (or hole) Wigner crystallization may exhibit interference between the triangular crystal sublattices, and inclusion of the resonance (hopping) between these two sublattices may cause blurring of the sharp transition to the insulator state, which seems compliant with observations (Fig. \ref{grafenrys4}).

\section{Conclusions}

In summary, we argue that, at fractional LL fillings, braid trajectories must be multi-looped, while those with lower number of loops (including single-looped) are excluded in the 2D case due to too short cyclotron radius. This unavoidable property of braids recovers Laughlin correlations in a natural way for 2D charged systems upon strong magnetic field and explains the structure of CFs both with flux-tubes or vortices. Classical cyclotron trajectories corresponding to braids with additional loops (as for fractional LL fillings) are also multi-looped and this property explains the true nature of effective models of  flux-tubes and vortices. Flux tubes attached to CFs do not actually exist and they only mimic additional cyclotron loops in the case of $\nu=\frac{1}{p}$ ($p$ odd). Out of the filling fraction $\nu=\frac{1}{p}$ ($p$ odd), the assumption on integer number of flux quanta attached to particles in order to create  CFs is, however, not clearly justified and only postulated in a heuristic manner. The introduced cyclotron braid approach allows for avoiding this postulate related to CF structure, including correction of mapping of FQHE onto IQHE and leading to recovery of LL filling hierarchy in a slightly modified version. Nevertheless, in order to rescue Jain's composite fermion structure with rigid integer flux quanta attached to each particle even outside $\nu =\frac{1}{p}$, one has to conjecture the relation between braid cyclotron picture and real cyclic movement of wave-packets represented particles and embracing by their orbits quantized fluxes. It leads to the eight-shape multi-looped quasiclassical trajectories in the case when the resultant Jain's field is oriented oppositely to the external field. The change of direction of cyclotron rotation would be measured in the experiment with cyclotron focusing of 2D carriers passing a narrow slot, by asymmetry of focusing to the left and to the right with respect to the source slot, when passing $\nu=\frac{1}{2}$ via changing the magnitude of the external field.

Unitary representations of cyclotron braids allow also for a self-consistent explanation of compressible states at fillings with even denominators. For example, $\nu =1/2$ metal Hall state corresponds to composite anyons with $p\alpha=3 \frac{1}{3} \pi=\pi$ signature of 1DUR of the $p=3$ cyclotron braid subgroup.

Recent experiments on suspended graphene  have indicated the crucial role of carrier mobility in the competition between Laughlin state and insulating state. Moreover, the fractional quantum Hall effect (FQHE) in graphene has been observed at  low carrier densities where the interaction was reduced due to particles dilution. This suggests that the interaction may not be as important  in the triggering of FQHE as expected based on the standard  formulation  of the composite fermion model.
Although   the real dynamics of quasiclassical wave packets is beyond the description ability  in the framework of a simplified 2D multiparticle charge system upon magnetic field, some general qualitative conclusions regarding the topological character can be drawn.  Via linkage of the classical cyclotron dynamics with quasiclassical wave packet trajectories, supported by a success of CF model with rigid flux quanta attached to particles, one can expect also multilooped structure of real quasiclassical wave packets dynamics.  Multilooped orbits have in 2D longer radii which strongly favor higher mobility of carriers and this has been confirmed experimentally in suspended graphene. The mobility is thus of primer significance for FQHE formation in competition with localized electron states like insulating Wigner-crystal-type-state. This picture seems to agree also with already predicted 
destabilization of Laughlin state in semiconductor heterostructures at $p>9$ in favor to Wigner crystal, when too many loops begin to be energetically unfavorable. Experiments with graphene have indicated  the triggering role of carrier mobility for FQHE state, which agrees with the topological explanation of CFs properties, thereby demonstrating that the standard concept of flux tubes generated by the interaction itself is insufficient.

%


\end{document}